\pdfoutput=1
\documentclass[10pt,aps,superscriptaddress,floatfix,preprint]{revtex4-2}   

\usepackage{gensymb}
\usepackage{color}
\usepackage[normalem]{ulem}
\usepackage{graphicx}
\usepackage{epstopdf}
\usepackage{dcolumn}
\usepackage{bm}
\usepackage{amsmath}
\usepackage{soul}
\usepackage[version=4]{mhchem}
\usepackage{hyperref}
\hypersetup{colorlinks,linkcolor=red,urlcolor=red,citecolor=blue}
\newcommand{\etal}{\emph{et al.}}
 
\newcommand{\be}{\begin{equation}}
\newcommand{\ee}{\end{equation}}
\newcommand{\bfig}{\begin{figure}}
\newcommand{\efig}{\end{figure}}

\newcommand{\GPS}{Gd$_2$PdSi$_3$}
\newcommand{\GG}{GdGa$_2$}
\newcommand{\GRS}{GdRu$_2$Si$_2$}

\begin{document}
\title{Triangular lattice magnet GdGa$_2$ with short-period spin cycloids\\ and possible skyrmion phases}

\author{Priya R. Baral}
\thanks{These authors contributed equally}
\affiliation{Department of Applied Physics and Quantum-Phase Electronics Center (QPEC), The University of Tokyo, Bunkyo-ku, Tokyo 113-8656, Japan}

\author{Nguyen Duy Khanh}
\thanks{These authors contributed equally}
\affiliation{Department of Applied Physics and Quantum-Phase Electronics Center (QPEC), The University of Tokyo, Bunkyo-ku, Tokyo 113-8656, Japan}
\affiliation{RIKEN Center for Emergent Matter Science (CEMS), Wako, Saitama 351-0198, Japan}

\author{Masaki Gen}
\affiliation{RIKEN Center for Emergent Matter Science (CEMS), Wako, Saitama 351-0198, Japan}
\affiliation{Institute for Solid State Physics, University of Tokyo, Kashiwa, Chiba 277-8581, Japan}

\author{Hajime Sagayama}
\affiliation{Institute of Materials Structure Science, High Energy Accelerator Research Organization, Tsukuba, Ibaraki 305-0801, Japan}

\author{Hironori Nakao}
\affiliation{Institute of Materials Structure Science, High Energy Accelerator Research Organization, Tsukuba, Ibaraki 305-0801, Japan}

\author{Taka-hisa Arima}
\affiliation{RIKEN Center for Emergent Matter Science (CEMS), Wako, Saitama 351-0198, Japan}
\affiliation{Department of Advanced Materials Science, The University of Tokyo, Kashiwa 277-8561, Japan}

\author{Yoshichika \={O}nuki}
\affiliation{RIKEN Center for Emergent Matter Science (CEMS), Wako, Saitama 351-0198, Japan}

\author{Yoshinori Tokura}
\affiliation{Department of Applied Physics and Quantum-Phase Electronics Center (QPEC), The University of Tokyo, Bunkyo-ku, Tokyo 113-8656, Japan}
\affiliation{RIKEN Center for Emergent Matter Science (CEMS), Wako, Saitama 351-0198, Japan}
\affiliation{Tokyo College, University of Tokyo, Tokyo 113-8656, Japan}

\author{Max Hirschberger}
\email{hirschberger@ap.t.u-tokyo.ac.jp}
\affiliation{Department of Applied Physics and Quantum-Phase Electronics Center (QPEC), The University of Tokyo, Bunkyo-ku, Tokyo 113-8656, Japan}
\affiliation{RIKEN Center for Emergent Matter Science (CEMS), Wako, Saitama 351-0198, Japan}

\maketitle

\begin{center}
\begin{large}
    Abstract
\end{large}
\end{center}
The two-dimensional triangular lattice (TAL) is a model system of magnetic frustration and competing interactions, where skyrmion spin vortices can be induced by a vertical magnetic field $B$. We target the binary compound \ce{GdGa2} with an undistorted TAL of Gd$^{3+}$ Heisenberg moments. At higher temperature ($T > 5$~K, $B = 0$, phase II), we reveal the cycloidal spin textures in \ce{GdGa2} via resonant elastic X-ray scattering (REXS). Further, a transition with strong magneto-elastic response occurs when cooling into the zero-field ground state ($T < 5$~K, phase I). We also report the thermodynamic phase boundaries of $B$-induced magnetic $A$-phases, which are suppressed by an in-plane magnetic field and which have enhanced resistivity due to the partial opening of a charge gap. In analogy to \ce{Gd2PdSi3} and \ce{GdRu2Si2}, these phases may represent a superposition of various cycloids, possibly a N{\'e}el skyrmion lattice. Our work lays the basis for further studies of the magnetic phase diagram of \ce{GdGa2}.

\section{Introduction}

The two-dimensional triangular lattice (TAL) is a model system of magnetic frustration and competing interactions. TAL magnets with antiferromagnetic nearest-neighbor interactions $J_1$ can host various ground states ranging from
commensurate 120$^\circ$ order to incommensurate helimagnetism~\cite{Gekht89}. Recently, the field-induced magnetic phases of TAL models have received special attention from theorists. Starting from a helimagnetic ground state, various numerical methods show complex field-induced magnetic textures, including skyrmion lattice phases~\cite{Okubo12, Hayami16, Hayami17, Kawamura24}. Such progress in theory has motivated experimental studies: in particular, spiral magnetism and a field-induced skyrmion phase was reported for \GPS~\cite{Frontzek09, Kurumaji19, Hirschberger20,hirschberger2020high}. Within the boundaries of the skyrmion lattice (SkL) phase, there appear dramatic responses related to the spin-winding number of magnetic moments in real space~\cite{Volovik87},
\begin{equation}
n_\mathrm{sk} = \frac{1}{4\pi}\int_\mathrm{m.u.c.} \bm{n}\cdot \left(\frac{\partial \bm{n}}{\partial x} \times \frac{\partial  \bm{n}}{\partial y}\right)\mathrm{d}x \mathrm{d}y
\end{equation}
where $\bm{n} = \bm{m} / \left|\bm{m}\right|$ is a normalized magnetization field and m.u.c. refers to the magnetic unit cell. Giant anomalies associated with $n_\mathrm{sk}$ have been reported in the Hall effect~\cite{Saha99, Kurumaji19}, in the thermoelectric Nernst effect~\cite{Hirschberger20}, in current-nonlinear dynamics~\cite{birch2024dynamic} and in the magneto-optical Kerr effect~\cite{Kato23}.

Here, we study the magnetic phase diagram of the TAL material \ce{GdGa2} in hexagonal space group $P6/mmm$ (\#191). Building on prior neutron diffraction studies~\cite{Barandiaran89,Hamasaki02a,Hamasaki04}, we use resonant elastic X-ray scattering (REXS) to study its cycloidal spin textures in two higher-temperature magnetic phases (phase II, phase A2). For the zero-field ground state (phase I), we present thermal expansion and magnetostriction measurements that demonstrate a crystallographic distortion with associated domain formation. We highlight the appearance of magnetic $A$-phases, induced by a magnetic field and highly sensitive to the tilting angle of the magnetic field with respect to the crystal axes. Drawing an analogy to established skyrmion host materials, we argue that this phase diagram in an incommensurate helimagnet is consistent with a topological SkL phase. We speculate that the centrosymmetric TAL material \ce{GdGa2} may host a N{\'e}el-type SkL with an extremely short magnetic period of only 0.54 nanometers.\\

\section{Magnetic phase diagram of $\mathrm{GdGa}_2$ with $B\parallel c$}

\begin{figure}[t]
  \begin{center}
  \includegraphics[clip, trim=0cm 0cm 0cm 0cm, width=0.7\linewidth]{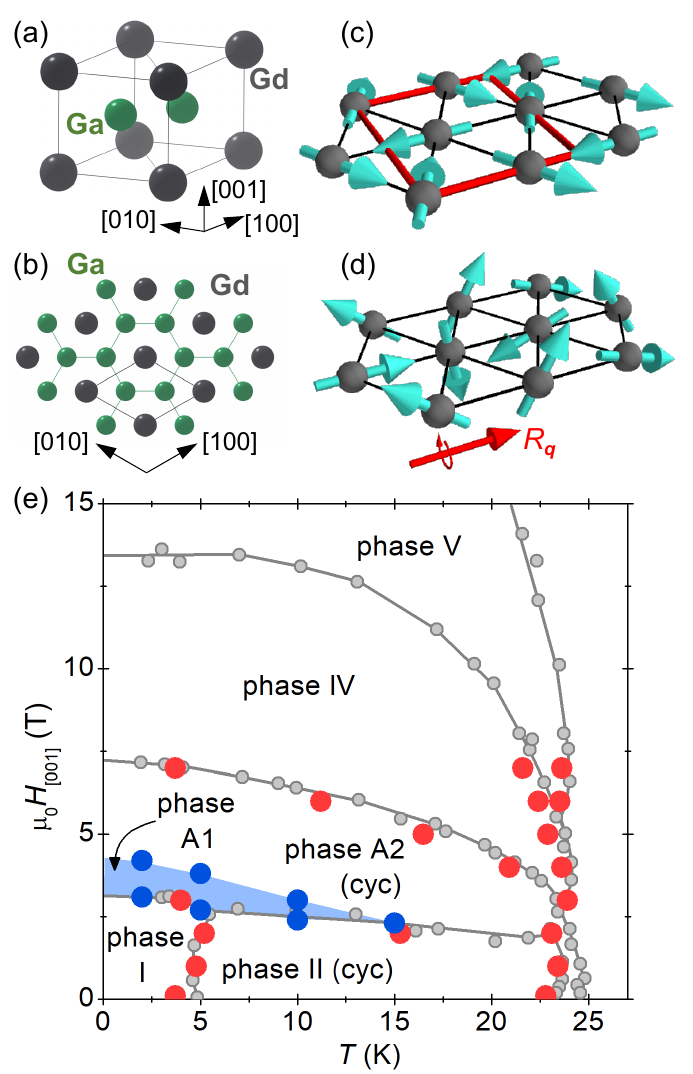}
   \caption[]{(color online). Magnetic order in the triangular lattice intermetallic \GG. (a,b) Crystal structure of \GG{} with triangular lattice of gadolinium (blue) and honeycomb lattice of gallium ions (green) visualized using VESTA~\cite{momma2008vesta}. The unit cell boundary is depicted by a black line. (c,d) Three-dimensional visualization of a triangular lattice with the well-known $120^\circ$ order (upper) and the cycloidal order reported here for phase II (lower). The red line indicates the boundary of the magnetic unit cell in (c). In (d), the rotation axis $R_{\bm{q}}$ is indicated (see text). (e) Magnetic phase diagram of \GG{} with magnetic field parallel to the $c$-axis. Grey dots adapted from Ref.~\cite{Hamasaki02b}, while red and blue circles are extracted from our magnetization data}.
   \label{Fig1}
  \end{center}
\end{figure}

Figure \ref{Fig1}a, and \ref{Fig1}b shows the hexagonal crystal structure of \GG{} with an undistorted TAL of gadolinium atoms and a honeycomb lattice of gallium atoms. In \GG{}, the triangular lattice motif is perfect, in contrast to the slightly distorted triangular lattice materials GdCu$_2$~\cite{Rotter00,karube2024giant} and Gd$_2$PdSi$_3$~\cite{Frontzek09}.
Competing interactions are generic to the high-symmetry, quasi two-dimensional TAL and a magnetic modulation vector $\bm{q}_\mathrm{cyc}= (0.39, 0.39, 0)$ was previously reported by neutron diffraction~\cite{Barandiaran89, Hamasaki02a, Hamasaki04}. 

Figure \ref{Fig1}c, and \ref{Fig1}d shows the main result of this article: the cycloidal magnetic texture of phase II, with $\bm{q}_\mathrm{cyc}$ incommensurate to the crystallographic structure in \ce{GdGa2}. We contrast this cycloid with the well-studied 120$^\circ$ antiferromagnetic order on the TAL [Fig.~\ref{Fig1}c]: This 120$^\circ$ state is commensurate with the TAL, while our $\bm{q}_\mathrm{cyc}$ corresponds to a slightly shorter magnetic period, with tilt angle 140$^{\circ}$ between neighboring magnetic moments. The 120$^\circ$ state is typically depicted as a planar arrangement of magnetic moments, as in Fig.~\ref{Fig1}c. However, the frustrated Heisenberg model does not have a preference for the spin plane, and a rotation by 90$^\circ$ around $\bm{q}_\mathrm{cyc}$ [$R_{\bm{q}}$ in Fig.~\ref{Fig1}d] transforms the cycloid into a planar arrangement.

The magnetic phase diagram of \GG{} reported in Fig. \ref{Fig1}e, based on magnetization and electric transport experiments, is consistent with prior work~\cite{Hamasaki02b}. One of our key results is the careful mapping of a novel phase pocket A1 by magnetization measurements (Supplementary Fig. \ref{SFig4})~\footnote{(Sec.~\ref{sec:SM}: Supplemental Material) See Supplementary Fig.~\ref{SFig4} for detailed magnetization measurements.}, which had so far drawn little attention~\cite{Gignoux95}.

\section{Experimental methods}
\label{sec:methods}
Single crystals of \GG{} grown by the Bridgman technique~\cite{Onuki23} (residual resistivity ratio RRR$\,\sim 16$) and by the floating zone technique (RRR$\,\sim 5$) were characterized by Laue X-ray diffraction (Fig. \ref{Fig2}b). Although samples made by both methods show comparable behavior, all data shown here was obtained from Bridgman samples. Some pieces were crushed to confirm the single-phase nature via powder x-ray diffraction. Many single crystals were found to show twin domains with similar $c$-axis alignment, but non-twinned pieces of lateral dimensions larger than $3\,$mm could be identified by careful inspection using an optical polarization microscope. We found that single crystals are heavily oxidized, with weakened magnetic anomalies (and no signature of phase I) after exposure to air for more than a week, so that all crystals were stored in vacuum. Magnetization measurements were carried out in a Quantum Design Magnetic Property Measurement System (MPMS) with rotator option, using carefully polished and aligned single crystal pieces. Magneto-transport measurements were carried out in a Quantum Design Physical Property Measurement System (PPMS) with a customized lock-in technique at excitation frequency $<20\,$Hz and excitation current $\sim5\,$mA. Thermal expansion and magnetostriction measurements along the $a$-axis were carried out by the fiber-Bragg-grating method using an optical sensing instrument (Hyperion si155, LUNA) using the PPMS.

For REXS experiments in reflection geometry, we mounted a plate-shaped single crystal with surface normal perpendicular to the $a^*$ crystal axis onto a rotation probe in the $8\,$T cryomagnet of beamline BL-3A of Photon Factory, KEK (Tsukuba, Japan). Figure \ref{Fig2}a illustrates the experimental configuration: Let the scattering plane be spanned by the incoming and outgoing beams $\bm{k}_i$ and $\bm{k}_f$, so that the momentum transfer $\bm{Q} = \bm{k}_i - \bm{k}_f$ lies within the $HK0$ plane. We further define the characteristic wavevector of the magnetic order in Fourier space as $\bm{q} = \bm{Q} - \bm{G}$, where $\bm{G}$ is the reciprocal lattice vector closest to $\bm{Q}$. In our discussion, $\bm{q}$, $\bm{Q}$, or $\bm{G}$ are given in reciprocal lattice units (r.l.u.). The incoming X-ray beam has its polarization in the scattering plane ($\pi$-polarization) and the energy of the beam is tuned to be in resonance with the Gd-$L_2$ absorption edge ($E_i=7.935\,$keV). Supplementary Fig. \ref{SFig1}~\footnote{(Sec.~\ref{sec:SM}: Supplemental Material) See Supplementary Fig.~\ref{SFig1} for technical details of resonant elastic X-ray scattering (REXS) experiments.} confirms the resonant scattering condition and in the temperature ($T$) scans of Supplementary Fig. \ref{SFig2}~\footnote{(Sec.~\ref{sec:SM}: Supplemental Material) See Supplementary Fig.~\ref{SFig2} for temperature dependence of resonant X-ray scattering (REXS) intensity in phases II ($B = 0$~T) and phase A2 ($B = 6$~T)}, magnetic reflections are absent above the N{\'e}el temperature $T > 25\,$K. Note that higher order reflections of the $\bm q_1$+$\bm q_2$-type could not be detected in the present experimental geometry, due to limited magnetic scattering intensity. 

For polarization analysis in REXS, we separate two components of the scattered beam at the detector: $\sigma^\prime$, perpendicular to the scattering plane, and $\pi^\prime$, within the scattering plane. Following Ref. \cite{Lovesey96}, the resonant scattering has a contribution $\sim \bm{m}_{\bm{q}} \cdot(\bm{\varepsilon}_f^* \times\bm{\varepsilon}_i)$, where $\bm{\varepsilon}_i$, $\bm{\varepsilon}_f$ are the x-ray electric field polarizations of the incoming and outgoing beam, respectively; $\bm{m}_{\bm{q}}$ is the Fourier component of the long-range ordered magnetic moment at momentum $\bm{q}$. Simplifying this expression, the two intensity components are $I_{\pi-\pi^\prime}(\bm{Q}) \propto \left(m_{\bm{q}}^z \sin 2\theta \right)^2$ and $I_{\pi-\sigma^\prime}(\bm{Q}) \propto \left(\bm{m}_{\bm{q}} \cdot \bm{k}_i \right)^2$, respectively. 

A systematic analysis of resonant scattering is carried out by considering the intensity ratio $R(\bm{Q}) = I_{\pi-\sigma^\prime}(\bm{Q}) / I_{\pi-\pi^\prime}(\bm{Q})$. For a N{\'e}el-type cycloid,
\begin{equation}
R \sin^2 (2\theta) = \left(\left|\bm{m}_{ip,\bm{q}} \right| / m_{\bm{q}}^z\right)^2 \cos^2\omega
\end{equation}
where $\omega$ is the angle $\angle(\bm{k}_i, \bm{q})$ when $\bm{k}_i$, $\bm{q}$ are in inverse \AA ngstroms (i.e., in a Cartesian frame). We define the parameter $E = \left(\left|\bm{m}_{ip,\bm{q}}\right| / m^z_{\bm{q}}\right)^2$ to quantify the elliptic distortion of the magnetic texture. In this article, the hexagonal crystal axes $a$, $a^{\ast}$, $b$, $b^{\ast}$, and $c$ are equivalently labeled as [100], (100), [010], (010), and [001] , respectively.

\section{X-ray scattering experiments}

\begin{figure*}[htb!]
  \begin{center}
    \includegraphics[clip, trim=0cm 0cm 0cm 0cm, width=0.99\linewidth]{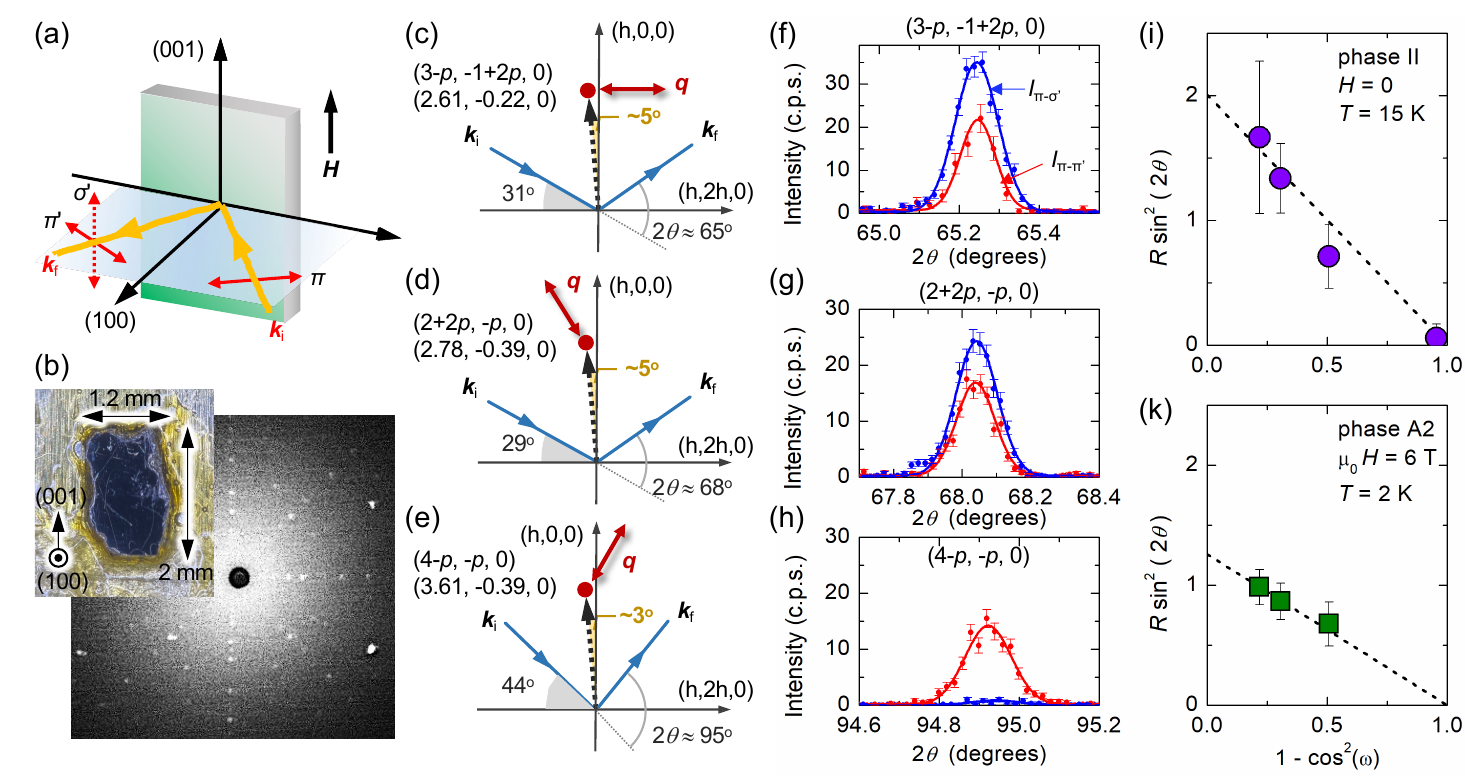}
    \caption[]{(color online). REXS of cycloidal magnetic orders in \GG{} at the Gd-$L_2$ edge. (a) Illustration of scattering geometry with polarization analysis. The incoming beam's polarization is in the scattering plane ($\pi$ polarization). (b) Optical image of the single crystal used for REXS mounted on an aluminum plate (upper) and Laue diffraction image with x-ray beam along the $(100)$ axis (lower). (c-e) Illustrations of incoming beam vector $\bm{k}_i$ and $\bm{q} = \bm{Q} - \bm{G}$, where $\bm{Q}$ and $\bm{G}$ are the momentum transfer and a reciprocal lattice vector, respectively. (f-h) Corresponding line scans with varying $\left|\bm{Q}\right|$, i.e., $\theta$-$2\theta$ scans. The REXS components $I_{\pi-\pi^\prime}$ (red) and $I_{\pi-\sigma^\prime}$ (blue) are shown for three reflections in phase II ($T = 15\,$K, $H =0$), with Poisson counting errors. (i, k) Polarization analysis from REXS for phases II, A2 of \GG. The dashed line corresponds to a model calculation for an elliptically distorted magnetic cycloid. Statistical errors of the fits to $I_{\pi-\pi^\prime}$ and $I_{\pi-\sigma^\prime}$ are propagated to $R$. See text for details.}
    \label{Fig2}
  \end{center}
\end{figure*}

Figure \ref{Fig2}c-h shows resonant scattering in phase II for three representative incommensurate reflections. The ordering vector $\bm{q}_\mathrm{cyc}$ of these reflections is in good agreement with prior neutron diffraction~\cite{Barandiaran89, Hamasaki02a, Hamasaki04}. As described in section \ref{sec:methods}, $I_{\pi-\pi^\prime}$ and $I_{\pi-\sigma^\prime}$ are sensitive to the magnetic texture components along the $c$-axis and within the scattering plane, respectively. The presence of $I_{\pi-\pi^\prime}$ at all three reflections in Fig. \ref{Fig2} therefore provides evidence for the $m_{\bm{q}}^z$ spin component. On the other hand, $I_{\pi-\sigma^\prime}$ varies significantly across the three reflections shown here and vanishes when $\bm{k}_i$ is perpendicular to $\bm{m}_{\bm{q}}$ at a given position $\bm{Q}$ in reciprocal space. We assume equal population of domains with $\bm{q}_{\mathrm{cyc},1} = (p, p, 0)$, $\bm{q}_{\mathrm{cyc},2} = (2p, -p, 0)$, and $\bm{q}_{\mathrm{cyc},3} = (p, -2p, 0)$, where $p = 0.39$. Then, the observed $I_{\pi-\sigma^\prime}$ intensities are in good agreement with the expected behavior of a spin cycloid. 
A survey of magnetic reflections for phase II is shown in Fig. \ref{Fig2}i; see section \ref{sec:methods} for the definition of $R$ and $\omega$. The data are consistent with slightly distorted cycloidal magnetic order (dashed line, ellipticity $E = 2.0(9)$)~\footnote{The data are inconsistent with spiral, fan, or sinusoidal textures in phase II. The latter two would have $I_{\pi-\sigma^\prime} = 0$; the spiral texture generates a line with a positive slope in this plot.}. Extending the REXS measurements to the field-induced magnetic phase A2 in Fig. \ref{Fig2}k, we find a less strongly distorted cycloidal magnetic pattern with $E =  1.2 (3)$. 

\section{Bulk measurements and thermodynamic phase diagram}

\begin{figure}[htb!]
  \begin{center}
	\includegraphics[clip, trim=0cm 0cm 0cm 0cm, width=0.8\linewidth]{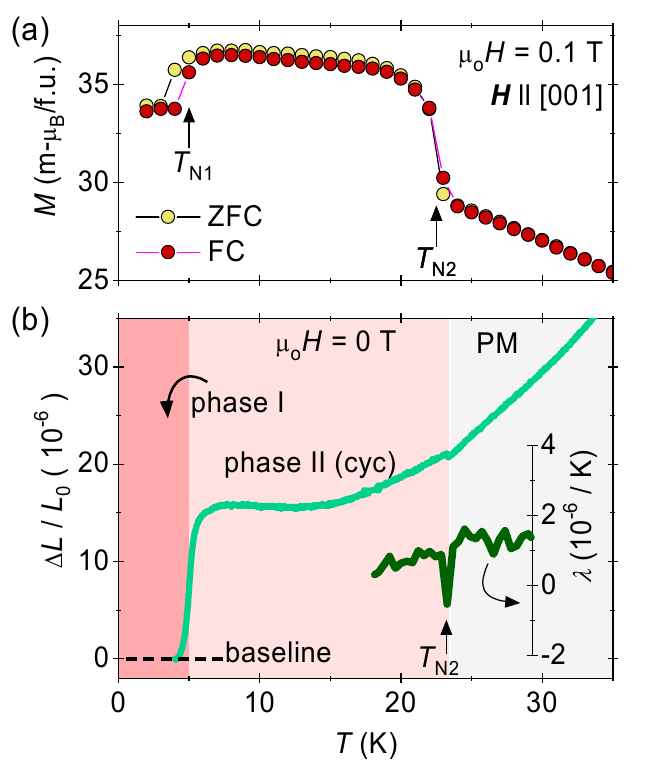}
    \caption[]{(color online). Bulk characterization of phase transitions in \GG{} in zero (or very low) magnetic field. (a) Magnetization $M(T)$ upon heating the sample, with two first-order transitions. ZFC and FC denote zero-field cooled and field cooled sample history, respectively. (b) Thermal expansion $L(T)$ along the $a$-axis upon heating, with a sharp anomaly upon entering phase II at $T_\mathrm{N1}$. In the thermal expansion coefficient $\lambda = \mathrm{d}\left(\Delta L / L_0\right) / \mathrm{d}T$, the transition at $T_\mathrm{N2}$ appears as a weak anomaly. }
    \label{Fig3}
  \end{center}
\end{figure}

In Fig. \ref{Fig3} we consider the low-temperature phase I via magnetization and thermal expansion experiments. There is a clear anomaly at $T_\mathrm{N1}$ in the thermal expansion $\Delta L(T)/L_0$, where $L_0$ is the length of the sample in zero magnetic field at $T = 4\,$K. This anomaly heralds significant spin-lattice coupling in this phase; domain formation associated with phase I is evident from hysteresis in the magnetostriction (Supplementary Fig. \ref{SFig3})~\footnote{(Sec.~\ref{sec:SM}: Supplemental Material) See Supplementary Fig.~\ref{SFig3} for magnetostriction data in \ce{GdGa2} at $T = 4$~K and $B || c$.}. 

In the context of phase I, we briefly review prior neutron diffraction experiments in \GG{}. A powder neutron study revealed magnetic intensity at $\bm{q}_\mathrm{cyc} = (0.39, 0.39, 0)$ in \GG{} at $T = 2\,$K, and no other reflections. The authors postulated a magnetic cycloid texture, but phases I was not distinguished from phase II~\cite{Barandiaran89}. Single crystal neutron diffraction later evidenced the formation of large domains in phase I~\cite{Hamasaki02a,Hamasaki04}: Hamasaki \etal{} reported the loss of magnetic intensity in phase I below $T_\mathrm{N1}$ for $\bm{Q}_1 = (0.39, 0.39, 0)$ and $\bm{Q}_2= (0.61, 0.61, 0)$; in this first experiment, the $c$-axis was vertical to the scattering plane~\cite{Hamasaki02a,Hamasaki04}. Upon rotating the crystal, they obtained rather different data, where $I(\bm{Q}_1)$, $I(\bm{Q}_2)$ increased below $T_\mathrm{N1}$~\cite{Hamasaki02a}. We propose that these two datasets can be consistently interpreted when considering the formation of large domains in phase I, and the effect of strain from the sample holder. The absence of at least one of three $\bm{q}$ domains can explain the suppression of some reflections below $T_\mathrm{N1}$. In summary, the neutron diffraction supports the notion that phase I has ordering vector $\bm{q} = (0.39, 0.39, 0)$ and large magnetic domains, likely associated with a structural symmetry lowering from hexagonal to orthorhombic or monoclinic symmetry. This scenario well explains the large anomaly in our thermal expansion data, and hysteresis in the magnetostriction.

\begin{figure}[htb!]
  \begin{center}
	\includegraphics[clip, trim=0cm 0cm 0cm 0cm, width=1.0\linewidth]{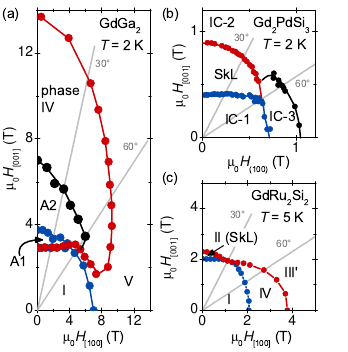}
    \caption[]{(color online). Tuning magnetic phases by a tilted magnetic field in \GG{} and related skyrmion lattice host materials. Phase boundaries of (a) \GG{} at $T = 2\,$K, where the area surrounded by red symbols has enhanced resistivity $\rho_{xx}$ in Fig. \ref{Fig5}. (b) Phase diagram for \GPS{} at $T = 2\,$K, and (c) tetragonal \GRS{} at $T = 5\,$K. The magnetic ordering vector $\bm{q}$ in the zero-field ground state is parallel to $a$, $a^*$, and $a$ in the three materials, respectively. Thus, in all three cases, magnetic field is rotated in the plane spanned by $\bm{q}$ and the $c$-axis. Panel (c) adapted from Ref.~\cite{Khanh22}.}
    \label{Fig4}
  \end{center}
\end{figure}

In Fig. \ref{Fig4}, we explore the phase diagram of \GG{} and related materials in a rotated magnetic field, at moderately low temperature. In the case of hexagonal \GPS{} and tetragonal \GRS{}, the ``outer'' phase boundary corresponds to the transition from spiral magnetism to a fan-like state. Between zero magnetic field and this threshold, there exist -- in each case -- magnetic phases that are highly sensitive to rotation symmetry breaking by an in-plane component of the magnetic field. These have been identified as skyrmion lattice (SkL) states, based on REXS, transport experiments, and/or real-space imaging experiments~\cite{Kurumaji19,Khanh20}. In hexagonal \ce{GdGa2}, phases A1 and A2 are most stable when the magnetic field is along the crystallographic $c$-axis, but are easily destabilized by a component $H_{[100]}$ in the basal plane. The skyrmion lattice in the Kagom\'{e} lattice material \ce{Gd3Ru4Al12} is also suppressed by an in-plane magnetic field~\cite{Hirschberger19,Hirschberger21b,Hirschberger24}, reflecting the generality of this experimental strategy for identifying centrosymmetric skyrmion materials where $\bm{q}$ is strongly pinned to preferred crystal axes.

\begin{figure}[htb!]
  \begin{center}
	\includegraphics[clip, trim=0cm 0cm 0cm 0cm, width=0.7\linewidth]{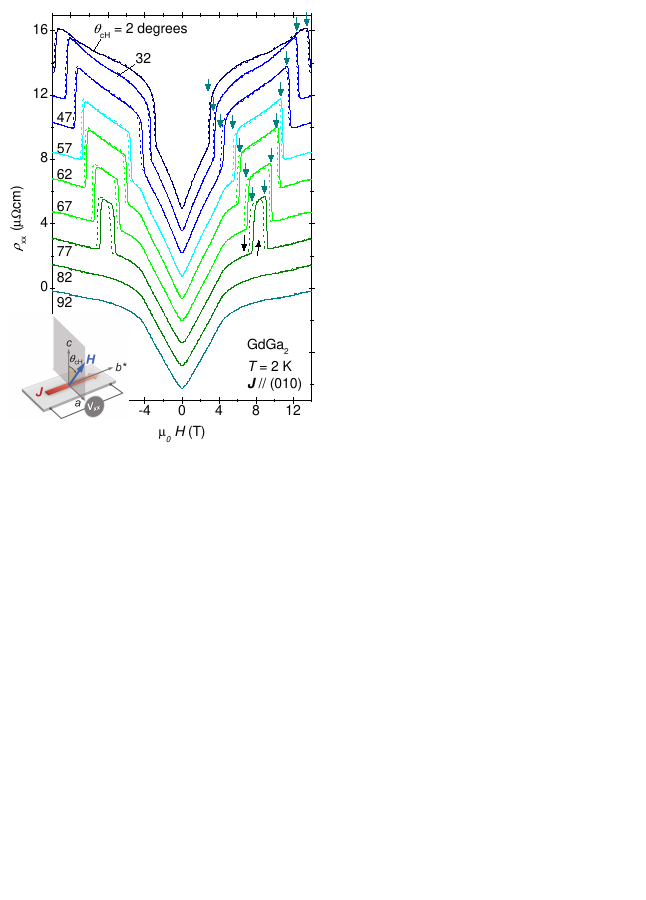}
    \caption[]{(color online). Transverse magnetoresistance $\rho_{xx}$ of \GG{} at $T = 2\,$K with current density $\bm{J}$ along the $b^*$ axis, i.e., at $90^\circ$ to the $a$-axis. The magnetic field is rotated in the $c$-$a$ plane and $\rho_{xx}$ is sharply enhanced in the regimes denoted as A1, A2, and IV in Fig. \ref{Fig4}. The lower boundary of phase A1, and the upper boundary of phase IV are indicated by vertical green arrows. The black arrows denote the direction of the magnetic field ramp for dashed ($\partial H /\partial t < 0$) and solid ($\partial H /\partial t > 0 $) lines. Curves are shifted downward in steps of $1.4\,\mathrm{\mu\Omega cm}$ as the angle $\theta_\mathrm{cH}$ increases.} 
    \label{Fig5}
  \end{center}
\end{figure}

The field-induced phases cause a dramatic change of the conducting properties of \GG: In Fig.~\ref{Fig5}, we show the magnetoresistance measured with magnetic field tilted at an angle $\theta_\mathrm{cH}$ away from the $c$-axis. Here, a sharp increase of $\rho_{xx}$ appears at the boundary between phases I and A1, remains roughly unchanged upon entering phase IV, and decreases again when transitioning into phase V. The anomaly in $\rho_{xx}$ neatly traces the phase boundaries illustrated in Fig \ref{Fig4} and is well-consistent with opening of a partial charge gap due to a multi-$\bm{q}$ magnetic order in an applied magnetic field~\cite{Wang20,Wang23}. A related anomaly in the Hall effect $\rho_{yx}$ was reported previously~\cite{Onuki23}.

\section{Discussion}
The magnetic order in phases II, A2 of \GG{} is incommensurate with $\bm{q}_\mathrm{cyc} = (0.39, 0.39, 0)$ and a rotation angle of $\sim 140^\circ$ between neighboring spin planes, in contrast  to commensurate $\bm{q}_{120^\circ} = (1/3, 1/3, 0)$ for the $120^\circ$ order on the TAL [Fig. \ref{Fig1}c]. To explain this incommensurability in the case of a metal, detailed electronic structure calculations are required; for example in \GPS~\cite{Inosov09, Nomoto20, Bouaziz22,Paddison22}, \GRS~\cite{Bouaziz22}, and NdAlSi~\cite{Bouaziz24}, states close to the Fermi energy $E_\mathrm{F}$ but also states far below $E_\mathrm{F}$ play an important role.

Although spin-orbit interaction, magnetic anisotropy, and spin-lattice coupling are expected to be weak for the $4f$ shell of Gd$^{3+}$ with total orbital angular momentum $L_{\mathrm{ang}} = 0$, we observe a remarkably large thermal expansion anomaly at the first-order transition between phases I and II. This effect may be related to an important contribution of anisotropic Gd-$5d$ electrons to the density of states at the Fermi energy~\cite{Kurumaji24}. 

Phases II and A2 both have cycloidal textures, are separated by a first-order phase transition, and the field-induced phase A2 is weak to an applied magnetic field in the TAL plane. It is thus likely that A2 is a linear superposition of several $\bm{q}_{\mathrm{cyc},i}$, for example a skyrmion lattice phase (triple-$\bm{q}$ state, $i = 1,2, 3$). As discussed in the introduction, such multi-$\bm{q}$ magnetic orders are expected from simulated annealing calculations both in the frustrated Heisenberg model when including interactions beyond $J_1$~\cite{Okubo12, Kawamura24} as well as in models of magnetic interactions from itinerant electrons~\cite{Hayami17}. Likewise the hitherto unreported phase A1, which only appears in a narrow window of temperature in magnetic field around $4\,$Tesla, may represent a complex superposition of various $\bm{q}_i$. 

\section{Conclusion}

Our REXS experiments confirm the magnetic ordering vector $\bm{q}_\mathrm{cyc} = (0.39, 0.39, 0)$ previously reported by neutron diffraction~\cite{Barandiaran89, Hamasaki02a, Hamasaki04}, and also reveal characteristic cycloidal textures of \GG{} consistent with the hypothesis of Barandiaran \etal{}~\cite{Barandiaran89}. Through REXS, we detect spin-cycloids in phases II and A2 in sharp contrast to related centrosymmetric intermetallics with spiral textures and Bloch skyrmion lattice phases~\cite{Kurumaji19, Hirschberger19, Khanh20, Khanh22, Hirschberger24}
. Phase I, the zero-field ground state, was not studied by REXS but shows a large anomaly in the thermal expansion $\Delta L / L_0$. Importantly, we observe unconventional magnetic $A$-phases in a window of temperature $T$ and magnetic field $B$, highly sensitive to a symmetry-breaking in-plane component of the magnetic field and with significantly enhanced resistivity $\rho_{xx}$. This is suggestive of a $C_3$ (or higher) rotation symmetry around the $c$-axis in phases A1 and A2 and a partial charge gap~\cite{Wang20,Wang23}. Further diffraction work is required to unambiguously determine the magnetic order in phases I, A1, and IV. \\

\noindent \textit{Acknowledgements}: This work was supported by JSPS KAKENHI Grants No. JP24H01607, JP23H05431, JP23K13069, JP23K13068, and JP22H04463. We acknowledge support from JST CREST Grant Number JPMJCR1874 and JPMJCR20T1 (Japan) and JST FOREST Grant No. JPMJFR2238 (Japan). The authors are grateful for support from the Murata Science Foundation, the Mizuho Foundation for the Promotion of Sciences, the Yamada Science Foundation, the Hattori Hokokai Foundation, the Iketani Science and Technology Foundation, the Mazda Foundation, the Casio Science Promotion Foundation, and the Takayanagi Foundation. P.R.B. acknowledges Swiss National Science Foundation (SNSF) Postdoc.Mobility grant P500PT\_217697 for financial assistance. Resonant X-ray scattering at Photon Factory (KEK) was carried out under proposal numbers 2022G551 and 2023G611.

\bibliography{GdGa2}

\newpage

\clearpage
\onecolumngrid
\begin{center}
\large{Supplemental Material}\\
\large{for}\\
\vspace{15pt}
{\LARGE``Triangular lattice magnet \ce{GdGa2} with short-period \\ spin cycloids and possible skyrmion phases''}\\
\vspace{15pt}
\large Priya R. Baral$^{1,\ddagger}$, Nguyen Duy Khanh$^{1,2,\ddagger}$, Masaki Gen$^{2,3}$, Hajime Sagayama$^{4}$, Hironori Nakao$^{4}$, Taka-hisa Arima$^{2,5}$, Yoshichika \={O}nuki$^{2}$, Yoshinori Tokura$^{1,2,6}$,
and Max Hirschberger$^{1,2,\ast}$\\
\vspace{10pt}
$^{1}$Department of Applied Physics and Quantum-Phase Electronics Center (QPEC), The University of Tokyo, Bunkyo-ku, Tokyo 113-8656, Japan\\
$^{2}$RIKEN Center for Emergent Matter Science (CEMS), Wako, Saitama 351-0198, Japan\\
$^{3}$Institute for Solid State Physics, University of Tokyo, Kashiwa, Chiba 277-8581, Japan\\
$^{4}$Institute of Materials Structure Science, High Energy Accelerator Research Organization, Tsukuba, Ibaraki 305-0801, Japan\\
$^{5}$Department of Advanced Materials Science, The University of Tokyo, Kashiwa 277-8561, Japan\\
$^{6}$Tokyo College, University of Tokyo, Tokyo 113-8656, Japan\\
\vspace{10pt}
$^{\ddagger}$These authors contributed equally\\
$^{\ast}$hirschberger@ap.t.u-tokyo.ac.jp\\

\end{center}
\clearpage

\renewcommand\thefigure{S\arabic{figure}} 
\setcounter{figure}{0}   

\renewcommand\thetable{S\arabic{table}} 
\setcounter{table}{0} 

\renewcommand\thesection{S\arabic{section}} 
\setcounter{section}{0}   

\section{Resonant elastic X-ray scattering (REXS) \& Bulk measurements}
\label{sec:SM}
\begin{figure*}[htb]
  \begin{center}
	\includegraphics[clip, trim=0cm 0cm 0cm 0cm, width=1.0\linewidth]{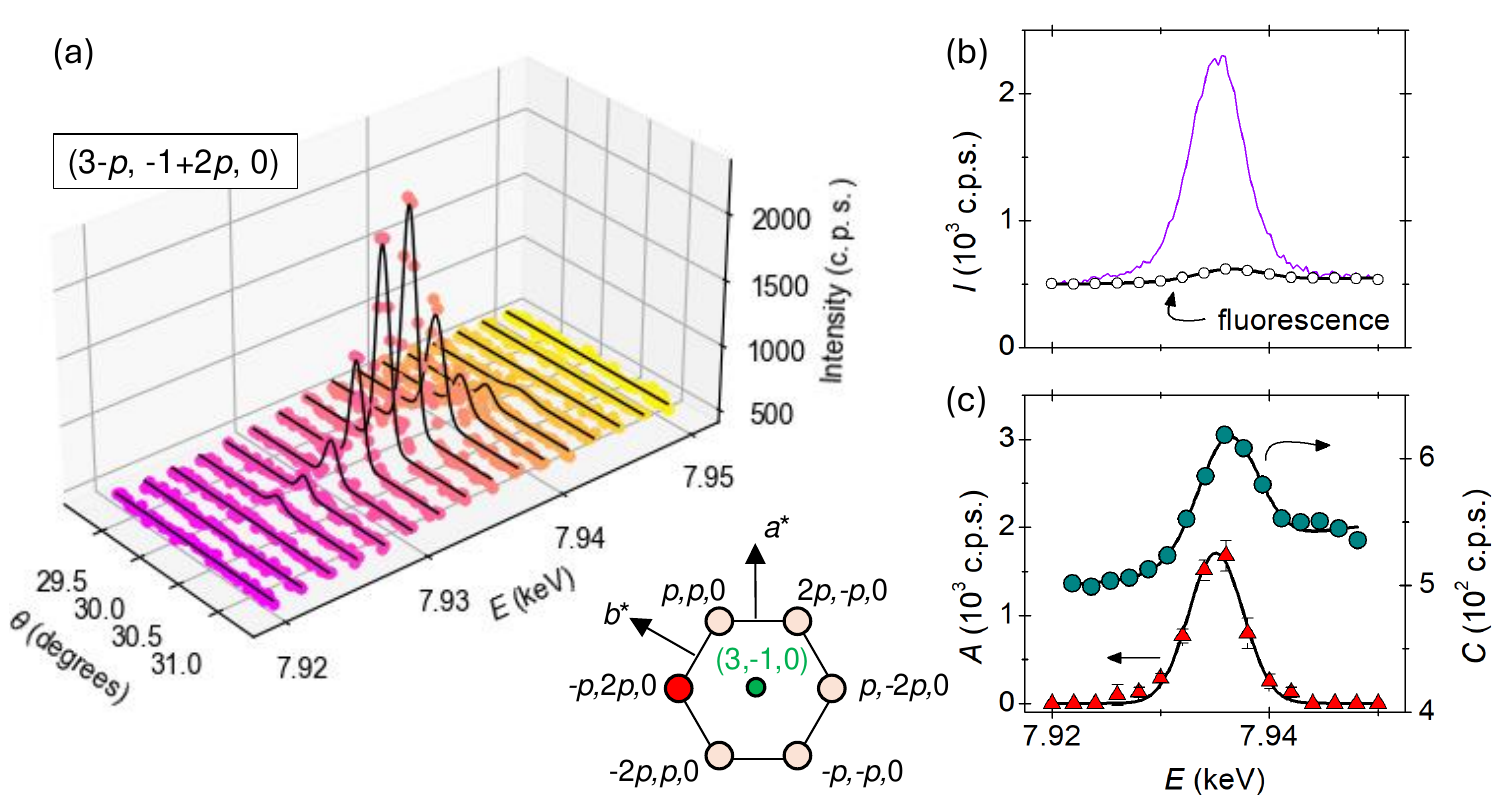} \caption[]{(color online). Resonant x-ray scattering in \GG{} without analyzer plate. (a) Rocking scans at various energies $E_i$ of the incoming x-ray beam, where $\theta$ is the scattering angle as defined by Bragg’s law. Clear resonant enhancement is observed around $E_i = 7.935\,$keV, corresponding to the Gd-$L_2$ absorption edge. Black lines: Gaussian fits $A\cdot\exp(-(\theta-\theta_0)/2\sigma^2)+C$ to the data. (b) Energy scan at fixed position in $\bm{Q}$-space (line) and the fluorescence signal (parameter $C$) from the Gaussian fits in panel (a) (symbols). (c) Left: Peak intensity $A$ from the Gaussian fits in panel (a), corresponding to the resonant part of the scattering signal. The line is a Gaussian fit to the energy profile. Right: Expanded view of fluorescence $C$ as in (b); the line is a guide to the eye. Lower inset: Illustration of magnetic satellites (red: reflection of interest) around the $(3,-1,0)$ fundamental reflection (green); $p = 0.39$ was found in phases II, A2.    }
    \label{SFig1}
  \end{center}
\end{figure*}

\begin{figure}[htb]
  \begin{center}
	\includegraphics[clip, trim=0cm 0cm 0cm 0cm, width=0.6\linewidth]{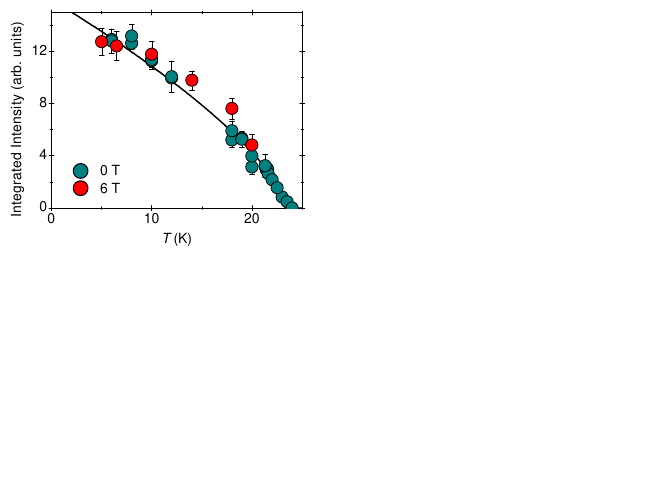}
    \caption[]{(color online). Temperature dependence of resonant x-ray scattering intensity in phases II ($B = 0\,$T) and phase A2 ($B = 6\,$T), without analyzer plate. We use the magnetic reflection depicted in the inset of Supplementary Fig. \ref{SFig1}. The line is a guide to the eye.}
    \label{SFig2}
  \end{center}
\end{figure}

\begin{figure}[htb]
  \begin{center}
	\includegraphics[clip, trim=0cm 0cm 0cm 0cm, width=0.6\linewidth]{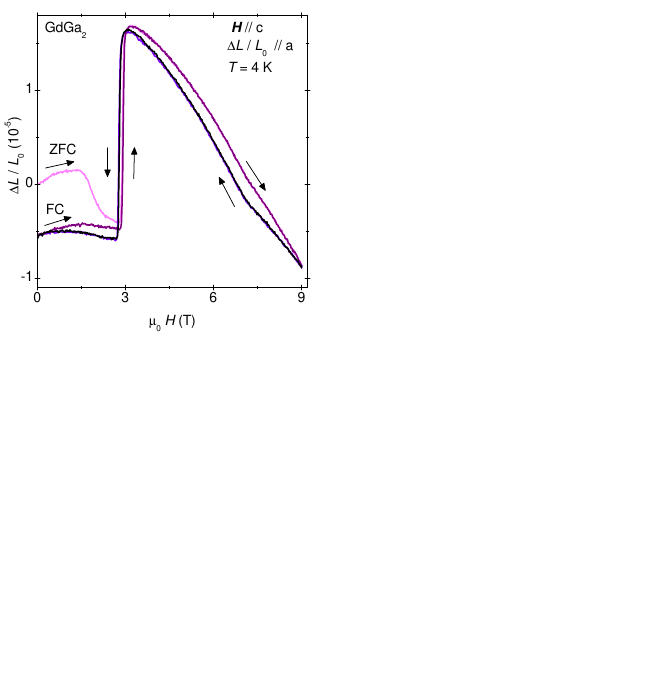}
    \caption[]{(color online). Magnetostriction in \GG{} at $T = 4\,$K and $B \parallel c$. Hysteresis around the transition between phase I ($B<2.8\,$T) and phase A1 ($B>2.8\,$T) indicates domain formation in phase I. We choose to measure the lattice expansion $\Delta L / L_0$ parallel to the $a$-axis, given that the modulation vector $\bm{q} = (0.39, 0.39, 0)$ is symmetry-equivalent to the $a$-axis in the hexagonal system.}
    \label{SFig3}
  \end{center}
\end{figure}

\begin{figure*}[htb]
  \begin{center}
	\includegraphics[clip, trim=0cm 0cm 0cm 0cm, width=0.9\linewidth]{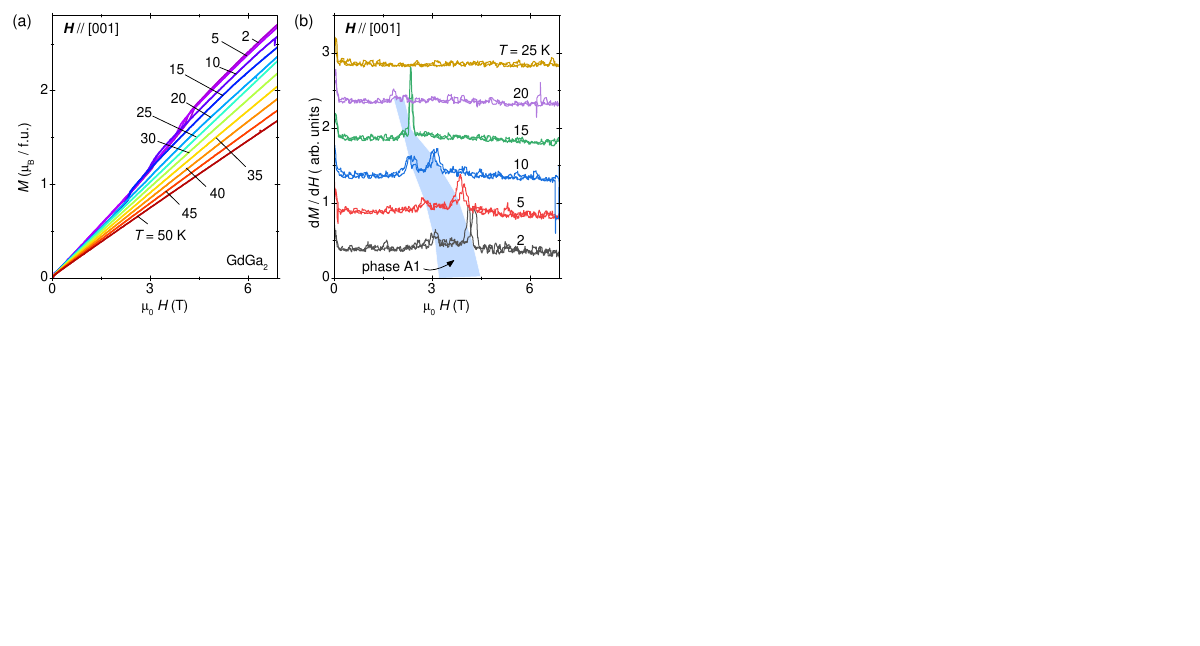}
    \caption[]{(color online). Magnetization $M$ and magnetic susceptibility $dM/dH$ for \GG{} when the magnetic field is parallel to the $c$-axis. We focus on the low-field regime, where a clear double-step transition indicates the presence of phase A1 (blue highlight) as discussed in the main text.}
    \label{SFig4}
  \end{center}
\end{figure*}

\end{document}